\documentclass
[twocolumn,english,prl,aps,floats,showpacs,superscriptaddress]{revtex4}%
\usepackage{graphicx}
\usepackage{epsfig,latexsym,amssymb,amsmath,amsbsy,graphics,graphicx}
\usepackage{dcolumn,bm,amsfonts}
\usepackage{babel}
\usepackage{amsmath}
\usepackage{amsfonts}
\usepackage{amssymb}%
\begin{document}
\title{Spin Liquid Behavior in Electronic Griffiths Phases}
\author{D. Tanaskovi\'{c}}
\affiliation{Department of Physics and National High Magnetic
Field Laboratory, Florida
State University, Tallahassee, Florida 32306}
\author{V. Dobrosavljevi\'{c}}
\affiliation{Department of Physics and National High Magnetic
Field Laboratory, Florida
State University, Tallahassee, Florida 32306}
\author{E. Miranda}
\affiliation{Instituto de F\'{\i}sica Gleb Wataghin, Unicamp,
Caixa Postal 6165, Campinas,
SP, CEP 13083-970, Brazil.}

\begin{abstract}
We examine the interplay of the Kondo effect and the RKKY interactions in
electronic Griffiths phases using extended dynamical mean-field theory
methods. We find that sub-Ohmic dissipation is generated for sufficiently
strong disorder, leading to the suppression of  Kondo screening on a finite
fraction of spins, and giving rise to universal spin-liquid behavior.

\end{abstract}

\pacs{71.27.+a, 75.20.Hr, 75.10.Nr}
\maketitle

Disorder induced non-Fermi liquid (NFL) behavior has remained an important
focus of study in heavy fermion systems \cite{experiments}. According to the
``magnetic Griffiths phase" scenario \cite{castroneto}, this behavior reflects
the formation of rare magnetic clusters with large susceptibilities, similarly
as in disordered insulating magnets \cite{griffiths}. Another approach focused
on the interplay of disorder and the Kondo effect far away from any magnetic
ordering \cite{mirandavladgabi1}. More recent work \cite{mirandavlad, darko}
has demonstrated that such ``electronic Griffiths phases" are a generic feature
of strongly correlated electronic systems with disorder. Neither picture,
however, seems satisfactory for the following key reason: in both scenarios,
the resulting NFL behavior is characterized by power law anomalies, with
non-universal, rapidly varying powers. In contrast, most experimental data seem
to show reasonably weak anomalies, close to marginal Fermi liquid behavior
\cite{experiments}.

Physically, it is clear what is missing from the theory. Similarly as magnetic
Griffiths phases, the electronic Griffiths phase is characterized
\cite{mirandavladgabi1,mirandavlad, darko} by a broad distribution $P\left(
T_{K}\right)  \sim\left(  T_{K}\right)  ^{\alpha-1}$ of local energy scales
(Kondo temperatures), with the exponent $\alpha\sim W^{-2}$ rapidly decreasing
with disorder $W$. At any given temperature, the local moments with
$T_{K}(i)<T$ remain unscreened. As disorder increases, the number of such
unscreened spins rapidly proliferates. Within the existing theory
\cite{mirandavladgabi1,mirandavlad, darko} these unscreened spins act
essentially as free local moments and provide a very large contribution to the
thermodynamic response. In a more realistic description, however, even the
Kondo-unscreened spins are \emph{not} completely free, since the metallic host
generates long-ranged Ruderman-Kittel-Kasuya-Yosida (RKKY) interactions even
between relatively distant spins. As the RKKY interaction has an oscillatory
character, the effective interactions between the (randomly located)
unscreened spins will be random in magnitude and sign. In this Letter, we use
an extended dynamical mean field formulation to examine the role of such RKKY
interactions within the electronic Griffiths phase scenario.

Our main results are as follows: (a) for disorder $W$ weaker than a critical
value $W_{c}$ we find Fermi liquid behavior, but for $W>W_{c}$ the Kondo
effect is suppressed on a finite fraction of spins, resulting in a spin-liquid
phase displaying (universal) marginal Fermi liquid behavior; (b) the spins
that remain screened are still characterized by a power law distribution of
(renormalized) energy scales $P\left(  T^{\ast}\right)  \sim\left(  T^{\ast
}\right)  ^{\alpha^{\ast}-1}$, but the exponent $\alpha^{\ast}$ acquires a
universal\emph{ }value\emph{ }$\alpha^{\ast}\approx1/2$ throughout the
spin-liquid phase; (c) the spin liquid phase is unstable to spin glass
ordering at the lowest temperatures, but we find robust marginal Fermi liquid
behavior in a broad temperature window above the freezing temperature.

We consider the disordered Kondo lattice model as given by the Hamiltonian
\begin{align}
H &  =-t\sum_{\langle ij\rangle\sigma}(c_{i\sigma}^{\dagger}c_{j\sigma
}^{\phantom{\dagger}}+\mbox{H.\, c.})+\sum_{i\sigma}v_{i}\,c_{i\sigma
}^{\dagger}c_{i\sigma}^{\phantom{\dagger}}\nonumber\\
&  +J_{K}\sum_{i}\mathbf{S}_{i}\cdot\mathbf{s}_{i}+\sum_{\langle ij\rangle
}J_{ij}\mathbf{S}_{i}\cdot\mathbf{S}_{j},\label{1}%
\end{align}
where the exchange couplings $J_{ij}$ between localized spins, and site
energies $v_{i}$ are distributed according to Gaussian distributions,
$P_{J}(J_{ij})\sim\exp(-J_{ij}^{2}/2J^{2})$ and $P_{W}(v_{i})\sim\exp
(-v_{i}^{2}/2W^{2})$ \cite{darko}. In this expression, $\mathbf{S}_{i}$ and
$\mathbf{s}_{i}=\frac{1}{2}\sum_{\alpha\beta}c_{i\alpha}^{\dagger
}\bm{\sigma}_{\alpha\beta}c_{i\beta}^{\phantom{\dagger}}$ represent a
localized spin and the conduction electron spin density at site $i$,
respectively. We concentrate on the paramagnetic phase. Applying the standard
procedure to average over disorder in the $J_{ij}$ couplings \cite{braymoore}
and taking the limit of infinite coordination $z\rightarrow\infty$
\cite{georgesreview}, the local effective action assumes the form
\begin{align}
\mathcal{A}_{j} &  =\sum_{\sigma}\int_{0}^{\beta}d\tau\int_{0}^{\beta}%
d\tau^{\prime}c_{j\sigma}^{\dagger}(\tau)[(\partial_{\tau}-\mu+v_{j}%
)\delta(\tau-\tau^{\prime})\nonumber\\
&  -t^{2}G_{c}(\tau-\tau^{\prime})]c_{j\sigma}^{\phantom{\dagger}}%
(\tau^{\prime})+J_{K}\int_{0}^{\beta}d\tau\mathbf{S}_{j}(\tau)\cdot
\mathbf{s}_{j}(\tau)\nonumber\\
&  -\frac{J^{2}}{2}\int_{0}^{\beta}d\tau\int_{0}^{\beta}d\tau^{\prime
}\chi (\tau-\tau^{\prime})\mathbf{S}_{j}(\tau)\cdot\mathbf{S}%
_{j}(\tau^{\prime}).\label{2}%
\end{align}
The local action of Eq. (2) describes the so-called Bose-Fermi Kondo model
\cite{sengupta,zhusi}, which describes a Kondo spin interacting with both a
fermionic bath of conduction electrons and a bosonic bath of spin
fluctuations. For a disordered Kondo lattice, we must
consider an ensemble of such
impurity models supplemented by following self-consistency conditions. The
bosonic spin bath $\chi(\tau)=\overline{\langle T_{\tau}%
\mathbf{S}_{j}(\tau)\cdot\mathbf{S}_{j}(0)\rangle_{\mathcal{A}_{j}}}=\int
dv_{j}P_{W}(v_{j})\langle T_{\tau}\mathbf{S}_{j}(\tau)\cdot\mathbf{S}%
_{j}(0)\rangle_{\mathcal{A}_{j}}$ and the conduction electron bath
$G_{c}(\tau)=\overline{G_{cj}(\tau)}=-\overline{\langle T_{\tau
}c_{j\sigma}^{\phantom{\dagger}}(\tau)c_{j\sigma}^{\dagger}(0)\rangle
_{\mathcal{A}_{j}}}$ are obtained by appropriate disorder averaging
\cite{darko}, and for simplicity we use a simple semicircular model density of
states for conduction electrons.

\textit{Destruction of the Kondo effect.} The presence of RKKY interactions
introduces a qualitative modification in the dynamics of the Kondo spins,
through the presence of a dissipative bosonic bath of spin fluctuations.
This behavior depends crucially on the precise spectral form of the bosonic
bath,
allowing for the destruction of the Kondo effect in the presence of sub-Ohmic
dissipation \cite{sengupta,zhusi}.  For a spectrum of the form
\[
\chi\left(  i\omega_{n}\right)  \sim\chi\left(  0\right)  -C\left\vert
\omega_{n}\right\vert ^{1-\varepsilon},
\]
Fermi liquid behavior is recovered for $\varepsilon=0$, but for
$\varepsilon>0$ (sub-Ohmic dissipation), and for sufficiently
small bare Kondo temperature $T_{K}$,  the spin decouples from the
conduction electrons. Within an electronic Griffiths phase,
however, the disordered Kondo lattice has a very broad
distribution of local Kondo temperatures $P\left(  T_{K}\right)
\sim\left(  T_{K}\right)  ^{\alpha-1}$. Therefore, for
$\varepsilon>0$ and arbitrarily weak coupling to the bosonic bath
(i.e. weak RKKY interaction), a fraction of the spins will
decouple.

To obtain a sufficient condition for decoupling, we examine the
stability of the Fermi liquid solution, by considering the limit
of infinitesimal RKKY interactions. To leading order we replace
$\chi(\tau)\longrightarrow\chi _{o}\left(  \tau\right)
\equiv\chi(\tau;J=0)$, \linebreak and the calculation reduces to
the {}\textquotedblleft bare model\textquotedblright\ of
Ref.~\cite{darko}.  The resulting bosonic bath, which is an
average over the site-dependent local dynamic spin susceptibility,
$\chi_{o}\left(  i\omega_{n}\right)  =\int dT_{K}P\left(
T_{K}\right) \chi \left(  T_{K},i\omega_{n}\right)$  , has a Fermi
liquid form in the presence of weak disorder. However, for
stronger randomness,
$W>W^{\ast}\approx\sqrt{t^{2}\rho_{c}J_{K}}/2$ corresponding to
$\alpha<2$ (here, $\rho_{c}$ is the density of states for
conduction electrons) \cite{darko} , the power law distribution of
energy scales within a Griffiths phase produces sub-Ohmic
dissipation, corresponding to $\varepsilon=2-\alpha>0$. Note that
the estimate based on the bare theory sets an \emph{upper bound}
for the true critical disorder strength, i.e.
$W_{c}<W^{\ast}=W_{nfl}/\sqrt{2}$ (here,
$W_{nfl}\approx\sqrt{t^{2}\rho _{c}J_{K}/2}$ is the threshold for
NFL behavior in the bare model \cite{darko}, corresponding to
$\alpha=1$). We emphasize that within the electronic Griffiths
phase, such decoupling emerges for $W>W_{c}$ even for arbitrarily
small $J$, in contrast to the clean case \cite{burdin} where much
stronger RKKY interactions ($J>J_{c}\approx10\,T_{K}$)
\cite{burdin10} are required to destroy the Kondo effect.

\textit{The spin liquid phase.} For finite $J$, the actual value of
$\varepsilon$ has to be self-consistently determined, as follows. For
$W>W_{c}$, the spins break up into two groups: the decoupled spins and those
that remain Kondo screened. Since the self-consistent bosonic bath function
$\chi(i\omega_{n})$ is an algebraic average over all spins, it is an additive
function of the contributions from each fluid
\begin{equation}
\chi\left(  i\omega_{n}\right)  =n\,\chi_{dc}\left(  i\omega_{n}\right)
+(1-n)\chi_{s}\left(  i\omega_{n}\right)  .\label{0sc1}%
\end{equation}
Here, $n$ is the fraction of spins in the decoupled phase. As we shall see,
the functions $\chi_{dc}\left(  i\omega_{n}\right)  $ and $\chi_{s}\left(
i\omega_{n}\right)  $ both have a singular, non-Fermi liquid form
characterized by exponents $\varepsilon_{dc}$ and $\varepsilon_{s}$,
respectively. Deferring for a moment the study of the critical region
(infinitesimally small $n$), we first examine the solution deep within the
spin liquid
phase. The first step in the self-consistent procedure is computing
$\varepsilon_{dc}$ and $\varepsilon_{s}$ for a given value of the bath
exponent $\varepsilon$. The spin autocorrelation function in the decoupled
phase assumes the form \cite{sengupta,zhusi} $\chi_{dc}(\tau)=\langle T_{\tau
}\mathbf{S}\left(  \tau\right)  \cdot\mathbf{S}\left(  0\right)  \rangle
\sim1/\tau^{\varepsilon}$, a result valid to all orders in $\varepsilon$
\cite{zhusi}. Since $\varepsilon_{dc}$ is defined by $\chi_{dc}(\tau
)\sim1/\tau^{2-\varepsilon_{dc}}$, we find
\begin{equation}
\varepsilon_{dc}(\varepsilon)=2-\varepsilon.\label{0sc3}%
\end{equation}
The non-analytic part of $\chi_{s}\left(  i\omega_{n}\right)  $ comes from the
spins with the smallest (renormalized) Kondo temperatures $T^{\ast}$
({}\textquotedblleft barely screened spins\textquotedblright)
\begin{equation}
\chi_{bs}\left(  i\omega_{n}\right)  =\int_{0}^{\Lambda}dT^{\ast}P\left(
T^{\ast}\right)  \chi_{bs}\left(  T^{\ast},i\omega_{n}\right)  .\label{0sc4}%
\end{equation}
Here $P\left(  T^{\ast}\right)  $ is the distribution of \emph{renormalized}
Kondo temperatures (local Fermi liquid coherence scales), and $\chi
_{bs}\left(  T^{\ast},i\omega_{n}\right)  $ is the local dynamic
susceptibility for a given $T^{\ast}$. Properties of the Bose-Fermi Kondo
model in the critical region of the  decoupling transition have been
extensively studied within renormalization group (RG) \cite{sengupta,zhusi}
and large-$N$ approaches \cite{burdin}, and we use these results to calculate
$\chi_{bs}$. In particular, $T^{\ast}\sim\left(  \delta J_{K}\right)  ^{\nu
}\sim\left(  \delta T_{K}\right)  ^{\nu}$, which gives $dT^{\ast}/dT_{K}%
\sim\left(  T^{\ast}\right)  ^{1-1/\nu}{}$. Therefore $P\left(
T^{\ast}\right)  =P\left[  T_{K}\left(  T^{\ast}\right) \right]
dT_{K}/dT^{\ast}\sim dT_{K}/dT^{\ast}\sim\left( T^{\ast }\right)
^{1/\nu-1}$. {}From scaling arguments \cite{sengupta,zhusi},
$\chi_{bs}\left(  T^{\ast },\omega\right) =\left(  T^{\ast}\right)
^{\eta-1}\phi\left(  \omega /T^{\ast}\right)  $, where $\eta$ is
the anomalous dimension, which is known to be exactly
$\varepsilon$ \cite{zhusi}. Performing the integration in
Eq.~(\ref{0sc4}), we find at low frequencies $\chi_{bs}\left(
i\omega _{n}\right) =\chi_{bs}(0)-C"\,\left\vert
\omega_{n}\right\vert ^{\eta +\frac{1}{\nu}-1}$, or, equivalently,
at large times $\chi_{bs}(\tau )\sim1/\tau^{\eta+\frac{1}{\nu}}$.
By definition, $\chi_{s}(\tau)\sim 1/\tau^{2-\varepsilon_{s}}$,
which gives
\begin{equation}
\varepsilon_{s}(\varepsilon)=2-\eta-1/\nu.\label{0sc6}%
\end{equation}
Since $\eta=\varepsilon$ and $\nu>0$ (as the relevant eigenvalue at the
unstable fixed point), Eqs.~(\ref{0sc3}) and (\ref{0sc6}) imply that
$\varepsilon_{dc}>\varepsilon_{s}$. Therefore, $\varepsilon
=\mbox{max}\{\varepsilon_{dc},\varepsilon_{s}\}=\varepsilon_{dc}$, and from
Eq.~(\ref{0sc3}) we find that the self-consistent bath is characterized by the
exponent $\varepsilon=1,$ as in the spin liquid model of Sachdev and Ye
\cite{sachdevye}, producing a logarithmic divergence of the average local
dynamic susceptibility.
Note that, in contrast to the bare ($J=0$) model of the
electronic Griffiths phase, the renormalized distribution $P\left(  T^{\ast
}\right)  $ of local energy scales now assumes a \emph{universal} form
characterized by an exponent $\alpha^{\ast}(\varepsilon)=1/\nu(\varepsilon
)\approx\varepsilon/2=1/2$ within the spin liquid phase. More work is needed
to determine the behavior of the uniform susceptibility, as well as
the behavior of the specific heat.

\textit{Transport in the spin liquid phase. }Although the renormalized Kondo
coupling scales to zero for the decoupled spins, the precise form of the RG
flows (scaling dimension of ``irrelevant operators") near the spin-liquid fixed
point still determines the finite frequency (or finite temperature)
corrections. To leading order, the contribution from decoupled spins scales as
$\rho(\omega)\sim\left[  J_{K}^{\ast}(\omega)\right]  ^{2}$, while
$J_{K}^{\ast}(\omega)\sim\omega^{1/\nu}$. To compute the appropriate exponent
at the spin-liquid fixed point we have used the  $\varepsilon$-expansion
approach of Ref.\cite{zhusi}, and we find  $\nu=2/\varepsilon+O(\varepsilon
^{3})$. From our self-consistent solution for the spin-liquid phase
($\varepsilon=1$), we obtain $1/\nu\approx1/2$, producing again a marginal
Fermi liquid correction to the resistivity $\delta\rho_{dc}(\omega)\sim\omega
$, or at $\omega=0$ and finite temperature
\[
\delta\rho_{dc}(T)\sim T.
\]

\textit{Numerical results.} As an illustration of our analytical predictions,
and to obtain quantitative results, we proceed to the numerical solution of
our equations in the large-$N$ limit \cite{sachdevye,burdin}. Introducing
site-dependent slave boson parameters $r_{j}$ and $\varepsilon_{fj}$, and
minimizing the local free energy, we come to the following saddle-point
equations \cite{burdin}%
\begin{align}
\frac{1}{\beta}\sum_{\omega_{n}}e^{i\omega_{n}0^{+}}G_{fj}(i\omega_{n}) &
=\frac{1}{2},\label{sp1}\\
\frac{1}{\beta}\sum_{\omega_{n}}G_{fj}(i\omega_{n})\Delta_{fj}(i\omega_{n}) &
=-\frac{1}{J_{K}}.\label{sp2}%
\end{align}
The local $f$-pseudo-fermion Green's function $G_{fj}(\tau)=-\langle T_{\tau
}f_{j\sigma}(\tau)f_{j\sigma}^{\dagger}(0)\rangle$, is given by $G_{fj}%
^{-1}\left(  i\omega_{n}\right)  =i\omega_{n}-\varepsilon_{fj}-\Sigma
_{j}\left(  i\omega_{n}\right)  -r_{j}^{2}\Delta_{fj}\left(  i\omega
_{n}\right)  $. The self-energy is equal to $\Sigma_{j}(\tau)=J^{2}%
\chi(\tau)G_{fj}(\tau)$, and $\Delta_{fj}^{-1}(i\omega_{n}%
)=i\omega_{n}+\mu-v_{j}-t^{2}G_{c}\left(  i\omega_{n}\right)  $.
Self-consistency requires  $\chi(\tau)=-\overline
{G_{fj}(\tau)G_{fj}(-\tau)}$, and $G_{c}(i\omega_{n})=\overline
{G_{cj}}(i\omega_{n})$, where $G_{cj}^{-1}\left(
i\omega_{n}\right) =\Delta_{fj}^{-1}\left( i\omega_{n}\right)
-r_{j}^{2}/\left[  i\omega _{n}-\varepsilon_{fj}-\Sigma_{j}\left(
i\omega_{n}\right) \right] .$\begin{figure}[ptb]
\begin{center}
\includegraphics[  width=3.3in,
keepaspectratio]{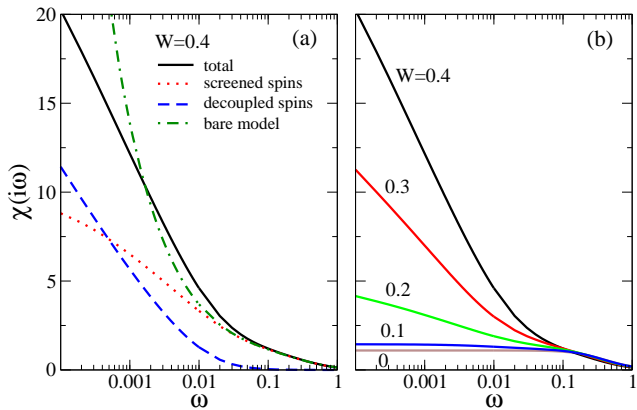}
\end{center}
\caption{Local dynamic magnetic susceptibility. (a) At low frequencies, the
decoupled spins (dashed line) provide the leading logarithmic dependence of
the total averaged susceptibility (full line). For the values of the
parameters used
($J_{K}=0.8$, $J=0.05$, $\mu=-0.1$ in units of the half bandwidth,
corresponding to $T_{K}\left(  v_{j}=0\right)  =0.1$ and $W_{c}\approx0.1$),
there are $n=8\%$ of decoupled spins at $W=0.4$. The bare model ($J=0$) leads
to a stronger non-universal power law singularity (dash-dotted line). (b)
$\chi(i\omega)$ for the disorder strength ranging from 0 to 0.4. }%
\label{fig1}%
\end{figure}

These equations were solved on the imaginary axis at $T=0$ using fast Fourier
transform methods. The total average local dynamic susceptibility
$\chi$ together with the
contributions coming from Kondo screened $\chi_{s}$ and decoupled spins
$\chi_{dc}$ is shown in Fig.~\ref{fig1}(a). At low frequencies, the
contribution from Kondo screened  spins saturates to a constant, while the
decoupled spins produce a logarithmic divergence. A comparison with the bare
model illustrates how the strong power law divergence of $\chi$ found for
$J=0$ is suppressed by the dynamical RKKY interactions. Fig.~\ref{fig1}(b)
shows how $\chi$ evolves with the change of disorder. Note that marginal Fermi
liquid behavior persists up to a crossover scale
$\omega_{sl}\sim0.1T_{K} \! \left(v_{j}=0\right)$
which has very weak dependence on the disorder strength.

\textit{Critical behavior.} Near the critical point the arguments which
followed Eq.~(\ref{0sc1}) have to be modified since the relative importance of
the various contributions to the average local susceptibility changes. First,
we concentrate on the contribution from the barely screened spins given by
Eq.~(\ref{0sc4}). As before, $P\left(  T^{\ast}\right)  =P\left[  T_{K}\left(
T^{\ast}\right)  \right]  dT_{K}/dT^{\ast}$, but close to the transition
$P\left(  T_{K}\right)  $ is small and cannot be replaced by a constant
prefactor of order $1$. Since $P\left(  T_{K}\right)  \approx P\left(
T_{Kc}\right)  \sim\left(  T_{Kc}\right)  ^{\alpha-1}$, we find $n=\int
_{0}^{T_{Kc}}dT_{K}P\left(  T_{K}\right)  \sim\left(  T_{Kc}\right)  ^{\alpha
}$, where $T_{Kc}$ is the bare Kondo temperature at the site energy $v_{c}$ at
which the spins start to decouple. Therefore $P\left(  T_{K}\right)  \sim
n^{(\alpha-1)/\alpha}$. From the bare model, we know that (for small $J$)
$\alpha\approx2$ near the critical point. Now we are in a position to write
down the general form of the total bosonic bath at low frequencies
\begin{equation}
\chi(i\omega_{n})=\chi_{o}-C_{1}\left\vert \omega_{n}\right\vert -C_{2}%
n^{1/2}\left\vert \omega_{n}\right\vert ^{\eta+\frac{1}{\nu}-1}-C_{3}%
\,n\,\ln\left\vert \omega_{n}\right\vert .\label{sc3}%
\end{equation}
\begin{figure}[ptb]
\begin{center}
\includegraphics[  width=2.9in,
keepaspectratio]{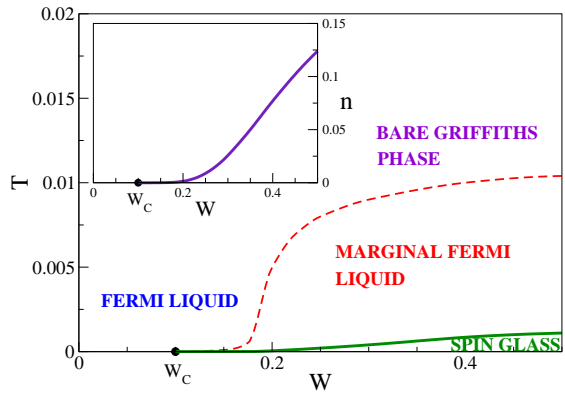}
\end{center}
\caption{Phase diagram obtained for the same values of parameters as in
Fig.~\ref{fig1}. The inset shows the fraction of decoupled spins as a function
of disorder. Note that the decoupling sets in already in the presence of
moderate disorder.
}%
\label{fig2}%
\end{figure}

The first two terms come from the well screened spins and have the
Fermi liquid form. The third term is due to the
{}\textquotedblleft barely screened\textquotedblright\ spins and
the last term is the contribution from the decoupled spins. The
crucial point is that the non-analytic term from the barely
screened spins, being proportional to $\sqrt{n}$, is much larger
than the logarithmic term due to the $n$ decoupled spins, except
at \emph{exponentially small} frequencies. Therefore, we can
neglect the last term in Eq.~(\ref{sc3}). Below the crossover
frequency $\omega^* \sim n^{(1/2)/[2-\eta-(1/\nu)]}$, the
non-analytic term in Eq.~(\ref{sc3}) is dominant and close to the
decoupling point, i.e. in the limit $\varepsilon_{fj}\rightarrow0$
and $r_{j}^{2}\rightarrow0$, $G_{f}$ assumes the form
$G_{f}(i\omega)\approx
-iC\,\mbox{sgn}\omega/|\omega|^{1-\varepsilon/2}$ for
$|\omega|<\omega^*$, and $G_{f}(i\omega)\approx -i/\omega$ for
$|\omega|>\omega^*$. The parameter $\varepsilon$ has to be
self-consistently determined from the equation
$1-\varepsilon=\eta+\frac{1}{\nu}-1$. Within the large-$N$ theory,
$\nu=2/\varepsilon$ which gives $\varepsilon=4/5$. Inserting this
expression into Eq.~(\ref{sp2}), we find the critical site energy
for decoupling $v_{c}\sim\sqrt{|\ln n|}$. Since the number of
decoupled spins is equal to $n=\int_{v_{c}}^{\infty}dvP_{W}(v)$,
we have a closed set of equations for $n(W)$, from which we find
that the number of decoupled spins is exponentially small in the
vicinity of the critical point
\begin{equation}
n\sim e^{-A/(W-W_{c})},\label{nsc5}%
\end{equation}
where $A$ is a positive constant. The numerical results for $n(W)$ are shown
in the inset of Fig.~\ref{fig2}. Interestingly, no precursors (vanishing
coherence scale of the Fermi liquid) arise as the critical point is approached
from the FL side, in contrast to what we have found by solving the same
equations in absence of disorder \cite{burdin}. This indicates a novel type of
quantum critical behavior that has a character of an essential singularity, a
feature that appears specific to quantum Griffiths phases.

\textit{Spin glass instability and phase diagram.} In this paper we have
concentrated on the paramagnetic solution of our model. However, the decoupled
spins can be expected to form a spin glass (SG) at low temperatures in the
presence of random inter-site interactions \cite{braymoore}. For a rigorous
treatment of the spin glass phase, one needs to go beyond the $N=\infty$
limit, but a rough  estimate of the temperature for SG ordering \cite{burdin}
may be obtained by using the large-$N$ approach as an approximate theory for
the considered  $N=2$ case. The spin glass instability criterion
\cite{braymoore}, as appropriately generalized to the case of additional site
randomness then reads%
\begin{equation}
\sqrt{\,\overline{\chi_{j}^{2}}\,}J/\sqrt{2}=1.\label{sg}%
\end{equation}
Fig.~\ref{fig2} represents a generic phase diagram of our model. For weak
disorder the system is in the Fermi liquid phase, while for $W>W_{c}$ the
marginal Fermi liquid phase emerges. The crossover temperature  (dashed line)
delimiting this regime can be estimated from the frequency up to which the
logarithmic behavior in $\chi(i\omega)$ is observed \cite{scaling}. The spin
glass phase, obtained from Eq.~(\ref{sg}), appears only at the lowest
temperatures, well below the marginal Fermi liquid boundary \cite{scaling}.
Interestingly, recent experiments have indeed found evidence of dynamical spin
freezing in the milliKelvin temperature range for some Kondo alloys
\cite{doug-sg}.

To summarize, we have introduced and solved a disordered Kondo lattice model
with random inter-site RKKY interactions. Our solution, valid within extended
dynamical mean-field theory,  illustrates how non-Ohmic dissipation arising
from inter-site RKKY interaction restores universality for non-Fermi liquid
behavior of electronic Griffiths phases. Although considerably different in
detail, this dissipative mechanism is reminiscent of the processes leading to
dynamical freezing of droplets within magnetic Griffiths phases \cite{noqcp},
suggesting a generic role of RKKY interactions in disordered heavy fermion
systems.

The authors thank Antoine Georges, Daniel Grempel, Dirk Morr, and Qimiao Si
for useful discussions. This work was supported by the NSF through grant
NSF-0234215 (V.D. and D.T.), FAPESP through grant 01/00719-8 (E.M.), CNPq
through grant 302535/02-0 (E.M.), FAEP through grant 0268/02 (E.M.), and the
NHMFL.


\begin{thebibliography}{99}

\bibitem {experiments}For a review of experiments on disordered heavy fermion
systems, see: G. R. Stewart, Rev. Mod. Phys. \textbf{73}, 797 (2001); D. E.
MacLaughlin \textit{et al.}, J. Phys, Condens. Matter \textbf{16},
S4479 (2004).

\bibitem {castroneto}A. H. Castro Neto \textit{et al.}, Phys. Rev. Lett.
\textbf{81}, 3531 (1998); A. H. Castro Neto and B. A. Jones, Phys. Rev. B
\textbf{62}, 14975 (2000).

\bibitem {griffiths}R. B. Griffiths, Phys. Rev. Lett. \textbf{23}, 17 (1969).

\bibitem {mirandavladgabi1}E. Miranda, V. Dobrosavljevi\'{c}, and G. Kotliar,
Phys. Rev. Lett. \textbf{78}, 290 (1997); J. Phys.: Condens. Matter
\textbf{8}, 9871 (1996).

\bibitem {mirandavlad}E. Miranda and V. Dobrosavljevi\'{c}, Phys. Rev. Lett.
\textbf{86}, 264 (2001).

\bibitem {darko}D. Tanaskovi\'{c}, E. Miranda, and V. Dobrosavljevi\'{c},
Phys. Rev. B \textbf{70}, 205108 (2004).

\bibitem {braymoore}A. J. Bray and M. A. Moore, J. Phys. C \textbf{13},
L655 (1980).

\bibitem {georgesreview}A. Georges \textit{et al.}, Rev. Mod. Phys.
\textbf{68}, 13 (1996).

\bibitem {sengupta}J. L. Smith and Q. Si, Europhys. Lett., {\bf 45}, 228
(1999); A. M. Sengupta, Phys. Rev. B \textbf{61}, 4041 (2000).

\bibitem {zhusi}L. Zhu and Q. Si, Phys. Rev. B \textbf{66}, 024426 (2002);
G. Zar\'{a}nd
and E. Demler, Phys. Rev. B \textbf{66}, 024427 (2002).

\bibitem {burdin}S. Burdin, D. R. Grempel, and A. Georges, Phys. Rev. B
\textbf{66}, 045111 (2002).

\bibitem {burdin10}We have numerically solved the equations of Ref.
\cite{burdin}, and found that $J_{c}\approx10\,T_{K}$, in agreement with an
improved analytical estimate (D. Grempel, private communication).

\bibitem {sachdevye}S. Sachdev and J. Ye, Phys. Rev. Lett. \textbf{70},
3339 (1993).

\bibitem {scaling}For quantitative results, we need to solve the large-$N$
equations at finite temperature. However, for the purpose of obtaining a rough
phase diagram, we have replaced temperature with frequency in $\chi_{j}\left(
T\right)  $, using the well-known $\omega/T$ scaling of the Bose-Fermi
Kondo model.

\bibitem {doug-sg}D. E. MacLaughlin \textit{et al}., Phys. Rev. Lett.
\textbf{87}, 066402 (2001).

\bibitem {noqcp} V. Dobrosavljevi\'{c} and E. Miranda, Phys. Rev. Lett.
\textbf{94}, 187203 (2005).
\end{thebibliography}
\end{document}